\documentclass[twoside,fleqn]{article}
\usepackage{amsfonts,espcrc2,latexsym}
 
\newcommand{\dalam}{\framebox[3mm]{}}
\newcommand{\cZ}{\cal{Z}}
\newcommand{\cS}{\cal{S}}
\newcommand{\cA}{\cal{S}}
\newcommand{\cN}{\cal{N}}
\newcommand{\cC}{\cal{C}}
\newcommand{\Ga}{\Gamma}
\newcommand{\dG}{\delta\Gamma}
\newcommand{\dr}{\delta\rho}
\newcommand{\ds}{\delta\sigma}
\newcommand{\dc}{\delta c}
\newcommand{\dA}{\delta A}
\newcommand{\pa}{\partial}
\newcommand{\ha}{\textstyle\frac{1}{2}}
\newcommand{\mn}{\mu\nu}
\newcommand{\Ge}{\Ga^{(1)}_{\mn}}
\newcommand{\Gz}{\Ga^{(2)}_{\mn}}
\newcommand{\Gn}{\Ga^{(0)}_{\mn}}
\newcommand{\emn}{\eta_{\mu\nu}}
\newcommand{\de}{\delta}
\newcommand{\Sc}{\scriptstyle} 
\newcommand{\ttbs}{\char'134}
\newcommand{\AmS}{{\protect\the\textfont2
  A\kern-.1667em\lower.5ex\hbox{M}\kern-.125emS}}
 
\title{The standard model in the on-shell scheme}
 \author{E. Kraus \address{Physikalisches Institut, Univ.\ Bonn,
        Nu\ss allee 12, D-53115 Bonn}
        \thanks{supported by Deutsche Forschungsgemeinschaft}
        and
        K. Sibold\address{Institut f.\ Theor.\ Physik, Univ.\ Leipzig,
        Augustus-Platz 10, D-04109 Leipzig}}

\begin{document}

\begin{titlepage}
%
\renewcommand{\thefootnote}{\fnsymbol{footnote}}
\begin{flushright}
BONN-TH/96-09\\
August 1996
\end{flushright}
\vspace{1cm}
 
\begin{center}
{\Large {\bf The standard model in the on-shell scheme}}\footnote{
To appear in the Proceedings of the Workshop ``QCD and QED in Higher
Order'', Rheinsberg, Germany (April 1996).} \\[4mm]
{\makebox[1cm]{  }       \\[2cm]
{\bf Elisabeth Kraus}\\ [3mm]
{\small\sl Physikalisches Institut, Universit\"at Bonn} \\
{\small\sl Nu\ss allee 12, D-53115 Bonn, Germany}} 
{and}\\[0.5cm]
{\bf Klaus Sibold }\\ [3mm]
{\small\sl Institut f\"ur Theoretische Physik, Universit\"at Leipzig} \\
{\small\sl Augustusplatz 10/11, D-04109 Leipzig, Germany} \\[0.5cm]
\vspace{2.0cm}
 
{\bf Abstract}
\end{center}
\begin{quote}
We outline the renormalization of the standard model to all orders of 
perturbation theory
in a way which does not make essential use of a specific subtraction scheme
 but is based
on the Slavnov-Taylor identity. Physical fields and parameters are used
 throughout the paper.
The Ward-identity for the global gauge invariance of the vertex
 functions is formulated.
As an application the Callan-Symanzik equation is derived.
\end{quote}
\vfill
\renewcommand{\thefootnote}{\arabic{footnote}}
\setcounter{footnote}{0}
\end{titlepage}
 
\begin{abstract}
We outline the renormalization of the standard model to all orders of 
perturbation theory
in a way which does not make essential use of a specific subtraction scheme
 but is based
on the Slavnov-Taylor identity. Physical fields and parameters are used
 throughout the paper.
The Ward-identity for the global gauge invariance of the vertex
 functions is formulated.
As an application the Callan-Symanzik equation is derived.

\end{abstract}
 
 
\maketitle

\section{Introduction}
The renormalization of the electroweak standard model is a well
studied subject (s.\cite{Hollik,JB1}). The one-loop approximation is
almost complete, two-loop calculations have been started.
 For the all order treatment however only \cite{BBBC}
 is available, where the necessary reasoning has been
performed in terms of the symmetric variables i.e.\ fields which would
occur without the spontaneous symmetry breaking.  But if one wants to
support the explicit calculations and e.g. formulate rigid invariance
as it is needed in practice then one has to renormalize in the
on-shell scheme which means formulate the renormalization in terms of
the physical fields.  Due to the $\gamma_5$-problem dimensional
regularization will not be invariant with respect to the symmetry
hence a formulation of the model is desirable which does not make
explicit use of a specific subtraction scheme, but relies only on
general properties like locality and power counting.  In the present
note we describe a few results of the respective analysis.
Specifically we write down the Slavnov-Taylor (ST) identity and discuss its
general solution in the classical approximation. As
 important applications we derive
then the rigid invariance and the Callan-Symanzik equation in the
one-loop approximation.  In order not to overload the paper we have
never written down the fermionic sector (i.e.  quarks, leptons and
their couplings) but mentioned when we made simplifying assumptions
(like CP-invariance) concerning them.

\section{The classical Lagrangean}
The standard model of electroweak interactions is a non-abelian gauge
theory with gauge group $SU(2) \times U(1)$. After spontaneous
symmetry breakdown one has to identify the physical fields $
W_\mu^\pm, Z_\mu, A_\mu$ and $H$ (for Higgs) in the bosonic sector,
quarks and leptons in the fermionic sector by diagonalizing the mass
matrices. At the same step one identifies the currents (charged,
neutral, electromagnetic) by constructing the charge eigenstates.
For the
bosonic sector we have the following classical approximation
\begin{eqnarray}
  \Gamma_{YM} & = &
  -\frac 1 4 \int G_a^{\mu\nu}\tilde{I}_{aa'}G_{\mu\nu a'}\\
  \Gamma_{scalar}&=&\int (D^\mu\Phi)^\dagger D_{\mu}\Phi - V(\Phi)\\
  V(\Phi)&\equiv&-\frac 1 8\frac{m_H^2}{M^2_W} g^2((\Phi^\dagger\Phi)^2
  \nonumber\\
  & &-v^2 (\Phi^\dagger\Phi)
  -\frac 1 4 v^4)\\
  \Phi &\equiv&\left(
    \begin{array}{c}
      \phi^+(x)\\
      1/\sqrt 2(v+H(x) + i\chi(x))
    \end{array}
  \right)
\end{eqnarray}
The field strength tensor and the covariant derivative have their usual form
\begin{eqnarray}
  G^{\mu\nu}_a & = & \partial ^\mu V_a^{\nu} - \partial^{\nu} V_a^\mu
  + g \tilde{I}_{aa'} 
  \hat{\varepsilon}_{a'bc} V^{\mu}_b V^{\nu}_c\\
  D_{\mu}\Phi &=& \partial_{\mu}\Phi - gi \frac{\hat{\tau}^a}{2} \Phi V_{\mu a}
\end{eqnarray}
We use the summation convention for the roman indices with values $+,
-, Z, A$ and have introduced convenient notations. The tensor
\begin{equation}
  \hat{\varepsilon}_{abc}  =  \left\{
    \begin{array} {ccc}
      \hat{\varepsilon}_{+-Z}&=&- i \cos\theta_W\\
      \hat{\varepsilon}_{+-A}& =&i \sin\theta_W 
    \end{array}\right.
\end{equation}
is completely antisymmetric and the matrices $\hat\tau_a\ \ (a=
+,-,Z,A)$ form a representation of $SU(2) \times U(1)$ according to
\begin{equation}
  \left[\frac{\hat{\tau}_a}2, \frac{\hat{\tau}_b}2\right]= 
  i \hat{\varepsilon}_{abc} 
  \tilde{I}_{cc'} \frac{\hat\tau_{c'}}2
\end{equation}
They are explicity given by $(\tau_i $, $i = 1,2,3$ the Pauli
matrices)
\begin{eqnarray}
  \hat{\tau}_+ \! 
= \hbox{$\frac 1{\sqrt{2}}$} (\tau_1 + i\tau_2) &\hspace{-3mm} &
\!
  \hat{\tau}_Z \! =\tau_3 \cos\theta_W -
  \mathbf {1}\frac{\sin^2\theta_W}{\cos\theta_W}
  \nonumber\\
  \hat{\tau}_-\!
 =\hbox{$\frac 1{\sqrt{2}}$} (\tau_1 - i\tau_2 ) &\hspace{-3mm} &
\!
  \hat{\tau}_A \!=-\tau_3 \sin\theta_W \! -\mathbf {1}\sin\theta_W 
\end{eqnarray}
The matrix $\tilde{I}_{aa'}$ guarantees the charge neutrality of the
classical action
\begin{eqnarray}
\tilde{I}_{+-}&=&\tilde{I}_{-+}=\tilde{I}_{ZZ}=\tilde{I}_{AA}=1\\
\tilde{I}_{ab}& =&0 \mathrm{\ else}
\end{eqnarray}
The value $ v=2 M_{W}/g $ for the shift parameter yields the desired
masses for the physical fields.  We may now turn to the global and
local symmetries of the action.  By construction we conserved the
electric charge and can obviously express this conservation law by
\begin{eqnarray} 
\int\!&i\bigl(&\hspace{-3mm} W_{\mu}^{+}\frac{\delta}{\delta W_{\mu}^{+}}-
    W_{\mu}^{-}\frac{\delta}{\delta W_{\mu}^{-}}\nonumber{}\\
 &+ &\hspace{-3mm} \Phi^{+}\frac{\delta}{\delta\Phi^{+}}-
    \Phi^{-} \frac{\delta}{\delta\Phi^{-}}\bigr)\Gamma_{GSW}
          = 0
\end{eqnarray}
\begin{equation}
\Gamma_{GSW}=\Gamma_{YM}+\Gamma_{scalar}
\end{equation}

According to Noethers theorem there exists an associated conserved
current which can be found via
\begin{equation}
  w_{em}\Gamma_{GSW}=-\partial_{\mu}j^{\mu}_{em} \qquad 
   \int w_{em} = {\cal{W}}_{em}
 \label{current}\nonumber{}\\
\end{equation}
with $w_{em}$
being the local charge operator. The explicit calculation yields
\begin{eqnarray}
  \partial_{\mu}j^{\mu}_{em} &=&
  -i \partial^{\mu}
  \Big(V_{+}^{\nu}G_{\mu\nu-}-
  V_{-}^{\nu}G_{\mu\nu+}\nonumber{}\\
 & & +  \phi^{-}(D_{\mu}\Phi)^{+}-
  \phi^{+}(D_{\mu}\Phi)^{-}
\Big)
\end{eqnarray}
which according to (\ref{current}) definitely {\it is} the
electromagnetic current but nevertheless does not have a simple
QED-like form because it contains non-abelian contributions in the
field strenght $G_{\mu\nu}$ and the covariant derivative $D_\mu \Phi$.
The final form of these relations is now obtained by verifying that
\begin{equation}
  \partial_{\mu}j^{\mu}_{em}=\frac 1{g \sin\theta_{W}}\partial^{\mu}
  \frac\delta{\delta A^{\mu}}\Gamma_{GSW}
\end{equation}
holds true and thus (\ref{current}) can be brought in the form of the
classical electromagnetic current identity
\begin{equation}
  \left( w_{em}+\frac1{g \sin \theta_{W}}
    \partial^{\mu}\frac\delta{\delta A^{\mu}}  \right)\Gamma_{GSW}=0
\label{classcurr}
\end{equation}
This functional form of Noethers theorem and local gauge invariance
permits to identify the coupling of the electromagnetic current to the
photon as the elementary charge $e$
\begin{equation}
  g \sin\theta_{W}=e
  \label{charge}
\end{equation}
It is noteworthy that the coupling $g$ of the abelian factor in $SU(2)
\times U(1)$ was not constrained by the algebra but fixed by the
assignment of the weak hypercharge and is now identified by
(\ref{charge}).

For the weak currents a similar reasoning is possible up to the point
where the spontaneous breakdown of the chiral $SU(2)_L$ symmetry
intervenes. So the global operator
\begin{eqnarray}
  {\cal{W}}_{cc}^{\pm}&\equiv& \sqrt 2 \int \hat\varepsilon_{ab\mp}V_{a}^{\mu}
  \tilde{I}_{bb'}\frac\delta{\delta V^{\mu}_{b'}}\nonumber\\
  & &+i\Phi^\dagger\frac{\tau_{\mp}}2
  \left(
    \begin{array}{c}
      \frac{\overrightarrow\delta}{\delta\phi^{-}}\\
      \frac1{\sqrt 2}\frac{\overrightarrow\delta}{\delta(H-i\chi)}\\
    \end{array}
  \right)\nonumber\\
  & &-i\left(\frac{\overleftarrow\delta}{\delta \phi^{+}},
  \frac1{\sqrt 2}\frac{\overleftarrow\delta}{\delta (H+i\chi)}\right)
  \frac{\tau_{\mp}}2 \Phi
\end{eqnarray}
when acting on $\Gamma_{GSW}$ will not produce zero but soft breaking
terms which correspond to the action of an inhomogeneous operator
\begin{equation}
  {\cal{W}}_{cc}^{\pm}\Gamma_{GSW}=-\frac ig \int \sqrt 2 M_{W}
  \frac{\delta \Gamma_{GSW}}{\delta \phi^{\mp}}
\end{equation}
The respective local equation reads
\begin{eqnarray}
  \lefteqn{w_{cc}^{\pm}\Gamma_{GSW}=-i\sqrt 2\frac{M_{W}}g 
    \frac{\delta\Gamma_{GSW}}{\delta\phi^{\mp}}
    -\partial_{\mu}j_{cc}^{\mu\pm}}\nonumber\\
  &=&-\sqrt 2  i\frac{M_{W}}g 
  \frac{\delta\Gamma_{GSW}}{\delta\phi^{\mp}}
  - \frac {\sqrt 2}g \partial^{\mu}
  \frac{\delta\Gamma_{GSW}}{\delta W^{\mu}_{\mp}}
  \label{locw}
\end{eqnarray}
and the bosonic contributions to the charged weak current comprise
also lower dimensional terms whose origin is the spontaneous breaking.
Analogously one proceeds for the neutral current.  The virtue of
(\ref{classcurr}) and (\ref{locw}) is that they exhibit the currents
and their (non-)conservation depending on their type.  For the
purposes of renormalization and compact writing one incorporates of
course the inhomogeneous differential operators into the l.h.s. and
defines
\begin{eqnarray}
  {\cal{W}}_{a}\equiv\int&\hspace{-3mm}\tilde{I}_{aa'}&\hspace{-3mm}
\Big( \hat \varepsilon_{bc'a'} V^{\mu}_{b}
  {\tilde I_{cc'}}\frac\delta{\delta V_{c'}^{\mu}}\nonumber\\
  & &+\Phi^\dagger\frac{i \hat\tau_{a'}}2
  \frac{\overrightarrow{\delta}}{\delta\Phi^\dagger}
  -\frac{\overleftarrow\delta}{\delta\Phi}\frac{i \hat\tau_{a'}}2 \Phi \Big)
\end{eqnarray}
as Ward identity operators satisfying the algebra
\begin{equation}
  [{\cal{W}}_{a},{\cal{W}}_{b}]=\hat\varepsilon_{abc}\tilde
  {I}_{cc'}{\cal{W}}_{c'}\quad a,b,c=+,-,Z,A
\end{equation}
The (broken) symmetry of $\Gamma_{GSW}$ under global transformations
is then expressed by the Ward identity
\begin{equation}
  {\cal{W}}_{a}\Gamma_{GSW}=0
\end{equation}
whereas the current identities assume the form
\begin{equation}
  \left( w_{a}+ \frac 1g \tilde I_{aa'} \partial^{\mu}
    \frac\delta{\delta V^{\mu}_{a'}} \right) \Gamma_{GSW}=0
  \end{equation}
  
\section{Gauge fixing, quantization and the Slavnor-Taylor identity}
The importance of the standard model originates from the fact that it is a 
renormalizable
field theory and thus permits consistent higher order calculations. 
A nessecary
prerequisite for this is gauge fixing which very often is done in the
terms of 
the so-called
$R_\xi$-gauges:
\begin{eqnarray}
\Gamma_{g.f.}&=&
  \int - \frac1{2\xi}F_{a}\tilde I_{aa'}F_{a'}\nonumber{}\\
F_+&\equiv&\partial_{\mu}W^{\mu}_{+}-iM_{W}\xi_{W}\phi_{+}\nonumber{}\\
F_Z&\equiv& \partial_{\mu}Z^{\mu}-M_{Z}\xi_{Z}{{\chi}}\nonumber{}\\
F_A&\equiv & \partial_{\mu}A^{\mu}
\label{rxigauge}
\end{eqnarray}
Applying the Ward identity operators ${\cal{W}} _a$ to $\Gamma_{g.f.}$
one finds that only ${\cal W} _A$ yields zero this being due to charge
neutrality. All other Ward identities are broken by the mass terms in
(\ref{rxigauge}).  This is a fortiori true for the 
local version of
which we give but one example:
\begin{eqnarray}
\lefteqn{\bigl(e w_{em}+\partial \frac\delta{\delta
    A}\bigr)\Gamma_{g.f.}=\frac {1} {\xi} \Box \partial A}\nonumber\\
\lefteqn{-\frac {ie}{2\xi}
 \partial^\mu [(\partial^\nu W_{\nu +}-i M_W \xi_W
    \phi_+) W_{\mu-} - \hbox{h.c.}]}
\end{eqnarray}
The terms in $W_\pm$ indicate that $\partial A$ is not a free field
hence unitarity cannot be deduced from the local Ward identity. In
order to remedy this situation one has to add the Faddeev-Popov ($\phi\pi$)
 fields
$c_a$, $\overline c_a$ $(a=+,-,Z,A)$ and to enlarge the gauge sector
by a ghost action in such a way that BRS invariance is achieved. The
BRS transformations of the physical fields read:
\begin{eqnarray}
\mathrm{s} V_{\mu a}&=&\partial _{\mu}c_a +g\hat \varepsilon_{abc}V_{\mu
  b}c_c\nonumber{}\\
\mathrm{s} c_{a}&=&- \frac g2 \hat \varepsilon_{abc}c_bc_c\nonumber{}\\
\mathrm{s} \bar c&=&-\frac 1 \xi F _a 
\end{eqnarray}
The complete action is invariant
\begin{equation}
  \mathrm{s} \Gamma_{\mathrm{cl}}
  = 0
\end{equation}
with 
\begin{equation}
\Gamma_{\mathrm{cl}} \equiv \Gamma_{\mathrm{YM}}
+\Gamma_{\mathrm{scalar}}+\Gamma_{\mathrm{g.f.}}+\Gamma_{\mathrm{ghost}} 
\end{equation}
BRS invariance turns out to be the relevant invariance for
quantization and renormalization because it fixes the interactions
amongst the unphysical fields in such a way that the complete action
is renormalizable and eventually the physical S-matrix unitary. The
proof of renormalizability and unitarity is based on a specific functional
form expressing the BRS invariance to all orders: the ST identity.
 The main reason for the formulation given below 
 is the questionable existence of
an invariant regularization for the standard model. 
One thus prefers
 not to rely on a very specific regularization scheme 
and to control the non-linear BRS transformations i.e.\ their
non-trivial renormalization by a suitable technical tool: external fields
coupled to the field variations.
\begin{eqnarray}
\Gamma_{\mathrm {ext}}&\hspace{-3mm}=&\hspace{-3mm}\int\!\rho^\mu_a
\tilde{I}_{aa'}\mathrm{s}V^\mu_{a'}
+\sigma_a \tilde{I}_{aa'}\mathrm{s}c_{a'}+Y_a \tilde{I}_{aa'} \mathrm{s}
\phi_{a'}
\end{eqnarray}
Since for the scalar fields (like for the vector fields) one cannot
stick to naive multiplet structure we label them also by the
indices $+,-,Z,A$:
\begin{equation}
(\phi_+,\phi_-,\phi_Z,\phi_A)\equiv (\phi_+,\phi_-,H,\chi)
\end{equation}

The other assignments of quantum numbers are such that $\Gamma_{\mathrm {ext}}$
has dimension four and is neutral with respect to electric and
$\phi\pi$ charges. The ST identity takes then the form
\begin{eqnarray}
  s (\Gamma)&\hspace{-3mm}\equiv &\hspace{-3mm}
  \int \frac{\delta \Gamma}{\delta \rho^{\mu}_a}
    \tilde{I}_{aa'}
    \frac{\delta \Gamma}{\delta  V_{\mu a'}}
+  \frac{\delta \Gamma}{\delta\sigma_a}
    \tilde{I}_{aa'}
    \frac{\delta \Gamma}{\delta c_{ a'}}\nonumber{}\\
& &\hskip5mm +B_a\frac{\delta \Gamma}{\delta \bar c_{a}}
+  \frac{\delta \Gamma}{\delta Y_a}
    \tilde{I}_{aa'}
    \frac{\delta \Gamma}{\delta  \phi_{ a'}}\quad = 0.
\label{sti}
\end{eqnarray}
and we have extended the classical action to
\begin{equation}
\Gamma=\Gamma_{\mathrm cl}+\Gamma^{(1)}+\Gamma^{(2)}+\ldots,
\end{equation}
the vertex functional ordered according to the number of closed
loops.  In writing down (\ref{sti}) we have slightly changed the gauge
fixing by introducing auxiliary fields $B_a$ of dimension $2$ and the
BRS transformations
\begin{equation}
\mathrm{s}\bar c_a=B_a \qquad \mathrm{s} B_a=0
\end{equation}
The respective gauge fixing reads
\begin{eqnarray}
\Gamma_{\mathrm {g.f.}}&= 
& \int\frac 12 \xi B_a \tilde{I}_{aa'}B_{a'}+ B_a \tilde{I}_{aa'} F_{a'}
\end{eqnarray}
The BRS transformations are then nilpotent  on all elementary fields and
$\Gamma_{\mathrm {g.f.}}+\Gamma_{\mathrm {ghost}} $
is a variation; this simplifies the algebraic analysis to be performed later on.

Quantization and consistent renormalization of the standard model
means to establish (\ref{sti}) to all orders of perturbation theory
for the 1PI Green functions collected in the functional $\Gamma$. 

The use of (\ref{sti}) proceeds in two steps: First one has to show that
the classical action (which is local) is the general solution of
(\ref{sti}) on which one has imposed a set of normalization conditions
i.e. one has disposed of a (finite) specific set of parameters. This
set must not be changed in higher orders otherwise one has no operator
interpretation in a fixed Fock space. Second one has to show that all
possible breakings of the ST identity arising in higher
orders can be removed by adjusting finitely many local counterterms.

For the purpose of the present paper we restrict ourselves to step one
and refer for step two to the literature \cite{BBBC}.  There it
has been shown (in terms of the "symmetric" variables) that the only
true obstruction to (\ref{sti}) is the Adler-Bardeen anomaly which
is absent in the standard model family by family due to the doublet
structure and colour multiplicity. 

In order to simplify for a first trial our task we assume that $CP$ is
conserved. This amounts to forbid mixing in the fermion families. We
choose the Higgs field to be even, the field $\chi$ to be odd under
$CP$.  Then we insert into (\ref{sti}) the most general
$\Gamma^{\mathrm{gen}}_{\mathrm{cl}}$ which is neutral with respect to
electric and $\phi\pi$ charge and has dimension less than or equal
to four. The outcome of the rather lengthy calculation can be
summarized as follows: The gauge fixing and ghost part of
$\Gamma^{\mathrm{gen}}_{\mathrm{cl}}$ is completely separated from the
rest. The vector and scalar field contributions are self-consistently
determined from the external field part, which fixes the
transformation properties of the fields. Input parameters are the
masses $M_W^2,M_Z^2,m_H^2$ of the vectors and Higgs respectively and
one coupling $g$. If one requires that these parameters do not vanish,
the action is uniquely determined up to (non-diagonal) field
redefinitions.  I.e. $\Gamma^{\mathrm{gen}}_{\mathrm{cl}}$ is obtained
from the standard form by performing the field transformations
\begin{equation}
  \begin{array}{cc}
    \begin{array}{rcl}
      V^a_{\mu}&\Rightarrow &\tilde z^V_{aa'} V^{\mu}_{a'}\\
      \phi^a_{\mu}&\Rightarrow&\tilde z^H_{aa'} \phi_{a'}\\
      c_{a}&\Rightarrow&\tilde z^G_{aa'} c_{a'}\\
    \end{array} &
    \begin{array}{rcl}
      \rho^a_{\mu}&\Rightarrow&\rho^{\mu}_{a'} (\tilde z^V)^{-1}_{a'a} \nonumber{}\\
      Y_a&\Rightarrow&Y_{a'} (\tilde z^H)^{-1}_{a'a} \nonumber{}\\
      \sigma_a&\Rightarrow&\sigma_{a'} (\tilde z^G)^{-1}_{a'a} 
    \end{array}
  \end{array}
\end{equation}
The general form of $\tilde z$ is the following 
\begin{equation}
\tilde {Z}=\left(
\begin{array}{cccc}
\tilde {z}_{++}&0&0&0\\
0&\tilde{z}_{--}&0&0\\
0&0&\tilde {z}_{ZZ}&\tilde {z}_{ZA}\\
0&0&\tilde {z}_{AZ}&\tilde {z}_{AA}\\
\end{array}
\right)
\end{equation}
The ST identity is invariant under these field
redefinitions.  CP-invariance yields $\tilde {z}^H_{ZA}=\tilde
{z}^H_{AZ}=0$, $\tilde {z}^V_{++}=\tilde {z}^V_{--}$, $ \tilde
{z}^H_{++}=\tilde {z}^H_{--}$, all real.

The parameters $\tilde z_{ab}$ have to be fixed by appropriate
normalization conditions. As an example we treat the bilinear vector
terms of the general action
\begin{eqnarray}
  \int  -\frac 14 (\partial^\mu V^\nu_a-\partial^\nu V^\mu_a)Z_{ab}^V \
  (\partial_{\mu} V_{\nu b}-\partial_\nu V_{\mu b})  \nonumber{}&&\\
  -\frac 12 V^\mu_a M_{ab}
  V_{\mu b}&&
\end{eqnarray}
where $Z^V_{ab}$ is a symmetric matrix defined by
\begin{equation}
Z^V_{ab}=\tilde z^T_{aa'}\tilde I_{a'b'} \tilde z_{b'b}
\end{equation}
This equation determines $\tilde z_{ab}$ only up to an orthogonal
matrix $O (\vartheta)$
\begin{equation}
O (\vartheta)=
\left(
\begin{array}{cccc}
  1&0&0&0\\
  0&1&0&0\\
  0&0&\cos \vartheta&\sin\vartheta\\
  0&0&-\sin\vartheta&\cos\vartheta
\end{array}
\right)
\end{equation}
hence it is convenient to write $\tilde z$ as
\begin{equation}
\tilde z=O (\vartheta) z
\end{equation}
with $z$ a symmetric matrix. After these preparations the mass matrix reads
\begin{equation}
(M_{ab})=M_Z^2 z (m_{ab})z
\end{equation}
where $(m_{ab})$ is symmetric with
$m_{+-}= \cos^2\theta_W$, $m_{ZZ}=\cos^2\theta_V$, $m_{ZA}=\sin\theta_V
\cos\theta_V$, $m_{AA}=\sin^2\theta_V$
and $\cos^2 \theta_W=M_W^2/M_Z^2$.
Continuing the example in the neutral sector we impose now the
normalization conditions \cite{AOKI}
\begin{equation}
\begin{array}{rcccl}
\partial_{p^2}\Gamma^T_{ZZ} (p^2=M_Z^2)&=&1&\Rightarrow&z_{ZZ}\\
\partial_{p^2}\Gamma^T_{AA} (p^2=0)&=&1&\Rightarrow&z_{AA}\\
\Gamma^T_{AZ} (p^2=0)&=&0&\Rightarrow&\theta_V\\
\Gamma^T_{AZ} (p^2=M_Z^2)&=&0&\Rightarrow&z_{AZ}\\
\Gamma^T_{ZZ} (p^2=M_Z^2)&=&0& &
\end{array}
\end{equation}

They separate the Z-boson and the photon on-shell and identify the
photon as the massless particle. As the explicit one-loop calculations
have shown one needs precisely these parameter redefinitions found
here in the general solution of the ST identity if one
wants to adjust the non-local mixing terms between $Z_{\mu}$ and
$A_{\mu}$.

\section{Ward-identities of global symmetries}
 
In section 2 we have indicated the global symmetries valid for the classical
action of the standard model, which are however violated in the course
of quantization by the mass terms of the gauge fixing. 
In higher orders the splitting up into gauge fixing and physical part 
becomes 
nontrivial,
 because the breaking can be inserted in any loop diagram with internal
vector and scalar lines, but still is soft, i.e.\
it vanishes for momenta larger than all masses.
 In order to formulate 
rigid symmetries in higher orders systematically and independently of the scheme
we couple the soft breaking to external non-propagating fields, so that the
symmetry is formally restored.
 Explicitly: First we enlarge the gauge fixing
by an external scalar doublet
\begin{equation}
\hat \Phi = {\hat \phi^ + \choose \frac 1 {\sqrt 2 }
(\hat v + \hat H + i \hat \chi )}
\end{equation} 
with the same quantum numbers as the scalar doublet \( \Phi \) and
a shift in its CP-even neutral component $\hat H $.
Then we write
\begin{eqnarray}
\Gamma_{{\rm g.f.}}&\hspace{-3mm}=&\hspace{-3mm}\int 
(\ha\xi B_a \tilde I _{ab} B_b +B_a \tilde I_{ab} \partial V_b \nonumber\\
&\hspace{-3mm} &\hspace{-3mm}-\frac {ie}{\sin \theta _W }
( \hat \Phi ^\dagger \frac {\hat \tau _a } 2  B_a  \Phi - 
 \Phi ^\dagger \frac {\hat\tau _a } 2  B_a  \hat \Phi)
\end{eqnarray} 
with $\hat\tau _a $ given in (8) and (9).

Equ.\ (46) is written in a manifestly invariant form \cite{DENNER}.
 It reduces to the 
original gauge fixing for vanishing external fields,
because it is the shift of the external Higgs, which produces the gauge
fixing of the quantum fields
\begin{equation} 
\hat v = \frac 2 e  \zeta \cos \theta_W \sin \theta _ W
 \quad\hbox{and} \quad \zeta _W =  \zeta _Z  = \zeta
\end{equation}
Enlarging ${\cal W} _a $ by the transformation of $\hat \Phi$
and $B_a $, the latter transforming in the same way as the vectors $V^\mu _a$,
$\Gamma_{\rm g.f.} $ is seen to be invariant. Finally we read off
from the BRS-transformations that the $\phi\pi$-ghosts transform also
like the vectors, and from the ST identity  we read off
the transformation of the
respective external fields. 
We thus arrive at
\begin{equation}
{\cal W} _a \Ga_{\rm cl} = 0
\end{equation}
\begin{eqnarray} 
{\cal W}_a =  \tilde I_{aa'} &\hspace{-2mm}
\int &\hspace{-3mm} \hat \epsilon _{bca'} V^\mu _b \tilde I _{cc'}
\frac{\delta}{\delta V^\mu _{c'}} \!
 + \{ \rho^\mu,\sigma, c, B, \bar c \} 
\\
& \hspace{-3mm}+&\hspace{-3mm}
  i \Phi ^\dagger \frac {\hat\tau _{a'} } 2  \frac{\overrightarrow 
\delta}{\delta \Phi^\dagger
}
 -  i \frac{\overleftarrow
\delta}{\delta \Phi} \frac {\hat\tau _{a'} } 2  \Phi  
+  \{ Y , \hat \Phi , q \} \nonumber
\end{eqnarray}
and
having formulated the symmetry as a Ward identity one again can interpret
$\Ga _{\rm cl}$ as the lowest order of the functional of 1PI Green functions
and deduce consequences for the 1-loop order 
 by using
 the action principle.
It tells us that the WI is at most  broken by local contributions, hence
all the non-local contributions sum up to symmetry.
It is clear that such identities constitute  important 
consistency checks in concrete calculations.
  
How to proceed for higher orders?
Neglecting for the moment the fermion sector with the parity violating
interactions we remain with the Higgs-vector model and
know that dimensional regularization is an invariant 
scheme.
Then  from the general classical action  one reads off that the
WI in its classical form is valid for the bare fields.
The renormalized Green functions satisfy therefore deformed WI which
have 1-loop corrections due to the field renormalizations (37).
 For the vector part this reads:
\begin{equation}
{\cal W}_a =  \tilde I_{aa'}
\int \!  V^\mu _b z_{bb'}\hat \epsilon _{b'c'a'} (\theta _W + 
\theta _V)\tilde I _{c'c}
z_{cd}
\frac{\delta}{\delta V^\mu _{d}}  
\end{equation}
 with $\theta_V = O(\hbar) $, $ z \equiv z^v = \tilde I + \delta z^{(1)}$
and $\delta z^{(1)}$ symmetric according to the definition (42).

For the complete theory this simple argument does  no longer hold and
 one has to carry out the proof
of rigid symmetry constructively by algebraic consistency.  The main point
is the observation that the deformed Ward operators (50) are derived from
an equivalence transformation and are the unique CP-odd solution of the
$SU(2) \times U(1) $ algebra (23).
Then  a deformed WI can be established including the fermion sector to 
all orders of perturbation theory.

\section{The Callan-Symanzik equation}

Finally as an application of 
rigid symmetry we derive the Callan-Symanzik
(CS) equation of the standard model in the spontaneously broken phase.
The CS equation describes the behaviour of the Green functions 
under dilatations. The hard breakings of dilatations which appear
due to the existence of asymptotic logarithms are the $\beta $-functions
and anomalous dimensions. These functions are constrained in one loop by the
global symmetry of the classical action.

The soft breaking of dilatations is in the classical approximation given by 
the mass terms, in higher orders it can be constructed along the lines of
\cite{EKS}. 
 
According to the action principle one has in 1-loop order
\begin{equation}
\underline m \partial _{\underline m} \Gamma =  [\Delta_m]_3^3
\cdot \Gamma + \alpha_i \Delta_i
\end{equation}
where $\underline m \partial _{\underline m} = M_Z \partial_ {M_Z}
+ M_W \partial_ {M_W}+m_H \partial_ {m_H} + \kappa \partial _\kappa
$ sums all massive parameters of the theory. 
$\Delta_i $ is a basis of all 4-dimensional polynomials in the fields
of the standard model. From BRS invariance one can deduce that these
local polymials can be written as differential operators acting on
the classical action. From 
algebraic consistency with rigid symmetry one finds that the hard
breakings are moreover globally symmetric:
\begin{equation}
{\cal W} _a \Delta _i \longrightarrow 0 \quad \hbox{for asymptotic momenta}
\end{equation}
This entails relations for the coefficient functions.
 The final
form of the 1-loop CS equation is given by:
\begin{eqnarray}
&\hspace{-3mm}
 &\hspace{-3mm}\Bigl(
\underline m \partial _{\underline m} +\beta_e e\pa_e +
\beta _{m_H} \pa _{m_H}  - 2 \hat \gamma _{\hat  \xi} \pa _{\hat \xi }
\nonumber \\
&\hspace{-3mm}&\hspace{-3mm}
+ \beta_{\scriptscriptstyle {M_W}}
\bigl( \sin \theta_ W \pa _{\scriptscriptstyle {\frac {M_W}{M_Z}}}
\nonumber \\ &\hspace{-3mm}&\hspace{1.5cm}
+  \int\! (
Z^\mu 
{\de}_{ A^\mu} - A^\mu {\de}_{ Z^\mu }  +\{B,\bar c,c \}) \bigl)  
\nonumber \\
&\hspace{-3mm} &\hspace{-3mm} - \gamma^V \int \!
\Bigl(V^\mu_a {\de}_ {V^\mu_a} -
B_a{\de}_{ B_a}-
 \bar c_a {\de}_{ \bar c_a} + 2 \xi \partial _\xi
\Bigr) \nonumber \\
&\hspace{-3mm}&\hspace{-3mm} - \hat\gamma^{V} \int \!\bigl((A^\mu + {\Sc 
\frac {\sin \theta_W}{\cos \theta _W }}
 Z^\mu)  ( {\de}_{ A^\mu} \!
 + {\Sc \frac {\sin \theta_W}{\cos \theta _W }}
{\de}_{ Z^\mu})\! -\! \{B ,\bar c \} \bigr) \nonumber \\
&\hspace{-3mm}&\hspace{-3mm} 
-\hat\gamma ^{g}
\int\! \bigl(c_ A + {\Sc \frac {\sin \theta_W}{\cos \theta _W }  }
 c_Z\bigr)\bigl({\de}_{ c_ A} + {\Sc \frac {\sin \theta_W}
{\cos \theta _W }}{\de}_{ c_Z} \bigr)   \nonumber \\
&\hspace{-3mm}&\hspace{-3mm} - \gamma ^g \int\!  c_a  {\de}_{ c_a} 
 - \gamma ^s \int\! \phi_a {\de}_{ \phi_a}  \Bigr) \Gamma
 =
 [\Delta_m]_3 ^3 \cdot \Gamma 
\end{eqnarray}

This form is valid for all tests with respect to propagating fields.
To simplify the writing the external fields have been put to zero.

There are two remarks concerning the CS equation: In the spontaneously
broken phase the CS equation contains the $\beta$-functions with respect
to the on-shell masses of the theory. A similar feature we have already
discussed in the $U(1)$-axial model with fermions \cite{EK}.

The anomalous dimensions $\hat \gamma$ appear in connection with the
abelian subgroup of the standard model. There we have to introduce
also the differentiation with respect to the abelian gauge parameter
$\hat \xi$; 
\begin{equation}
\int \hat \xi \bigl( 
\frac {\sin \theta _W }{\cos {\theta } _W } B_Z + B_A \bigr)^2
\end{equation}
This gauge fixing contains non-diagonal contributions and is generally
taken to be zero in lowest order. The fact, that it has to be
introduced into the CS equation with a nonvanishing coefficient
$\hat \gamma _{\hat \xi} =( \beta _e - 2 \beta_{\scriptscriptstyle {M_W}}
\frac {\sin \theta_W}{\cos \theta _W } )  \hat \xi + \hat  \gamma^V \xi $
 shows
that the choice $\hat \xi = 0 $ is not stable under renormalization.
 
Finally we want to mention that there is a relation between the 
$\beta$-functions and anomalous dimensions due to the abelian
subgroup:
\begin{equation}
\beta _e = \beta _{M_W} \frac {\sin \theta _W }{\cos \theta _W} +
\gamma ^V + \hat \gamma ^V \frac 1 {\cos^2 \theta _W}
\end{equation}

\section{Conclusions}
The ST identity governs the construction of gauge theories. It defines
the model in question once the symmetry group and the multiplets are given; it then
determines the rigid invariance which embodies the remaining physical symmetry.
Within the 
standard model
 we have sketched as an example how the general solution of the classical
approximation exhibits the free parameters
 which have to be fixed by normalization 
conditions or can partly be constrained by the rigid invariance. The expression
of the rigid symmetry in terms of a Ward identity turns out to be non-trivial.
The effect of the classical symmetry to the one-loop approximation and
the CS equation has been indicated.

\end{document}